%% file: main.tex
\documentclass[sigplan,screen,nonacm]{acmart}
\AtBeginDocument{%
  \providecommand\BibTeX{{%
    \normalfont B\kern-0.5em{\scshape i\kern-0.25em b}\kern-0.8em\TeX}}}

\usepackage{makecell}
\usepackage{listings}
\usepackage{color}
\definecolor{codegreen}{rgb}{0,0.6,0}
\definecolor{codegray}{rgb}{0.5,0.5,0.5}
\definecolor{codepurple}{rgb}{0.58,0,0.82}
\definecolor{backcolour}{rgb}{0.95,0.95,0.92}
\lstdefinelanguage{WebAssembly}{
  sensitive=true,
  otherkeywords={},
  morekeywords=[1]{i32,f32,i64,f64},
  keywordstyle={[1]\color{violet}},
  morekeywords=[2]{0},
  keywordstyle={[2]\color{violet}},
  morekeywords=[3]{add,const}
  keywordstyle={[3]\color{bluemunsell}},
  morekeywords=[4]{},
  keywordstyle={[4]\color{candypink}},
  morekeywords=[5]{module, func, param, result, global, get_global, mut, set_global, export, import, memory, data, get_local, set_local, elem, table, call,call_indirect, type},
  keywordstyle={[5]\color{blue}},
  morekeywords=[6]{=,;},
  keywordstyle={[6]\color{britishracinggreen}},
  morekeywords=[7]{(,),[,],.},
  keywordstyle={[7]\color{black}},
  numberstyle=\tiny\color{black},
  rulecolor=\color{black},
  morecomment=**[l][\itshape\color{greencomments}]{;;},
}

\lstdefinestyle{mystyle}{
    backgroundcolor=\color{backcolour},   
    commentstyle=\color{codegreen},
    keywordstyle=\color{magenta},
    numberstyle=\tiny\color{codegray},
    stringstyle=\color{codepurple},
    basicstyle=\ttfamily\footnotesize,
    breakatwhitespace=false,         
    breaklines=true,                 
    captionpos=b,                    
    keepspaces=true,                 
    numbers=left,                    
    numbersep=5pt,                  
    showspaces=false,                
    showstringspaces=false,
    showtabs=false,                  
    tabsize=2
}

\begin{document}

\title{Defending Buffer Overflows in WebAssembly: \\A Transpiler Approach}


\author{Weiqi Feng}
\authornotemark[1]
\affiliation{%
 \institution{Harvard University}
 \city{Cambridge}
 \state{MA}
  \country{USA}}
\email{wfeng@g.harvard.edu}



\renewcommand{\shortauthors}{Chen, Feng, Lao, and Li}

\newcommand{\lam}[1]{\textbf{\color{orange}lam: #1}}
\newcommand{\vic}[1]{\textbf{\color{red}vic: #1}}
\newcommand{\minghao}[1]{\textbf{\color{blue}minghao: #1}}
\newcommand{\codesm}[1]{\texttt{\small #1}}

\input{sections/abstract}
\maketitle

\input{sections/introduction}
\input{sections/background}
\input{sections/implementation}
\input{sections/evaluation}
\input{sections/limitation}
\input{sections/lesson}
\bibliographystyle{ACM-Reference-Format}
\bibliography{sample-base}
\appendix
\input{sections/appendix}









\end{document}

%% file: sections/abstract.tex
\begin{abstract}
  WebAssembly is quickly becoming a popular compilation target for a variety of code. However, vulnerabilities in the source languages translate to vulnerabilities to the WebAssembly binaries. This work proposes a methodology and a WebAssembly transpiler to prevent buffer overflows in the unmanaged memory of the WebAssembly runtime. The transpiler accepts a WebAssembly binary and adds stack canaries and Address space layout randomization (ASLR) to protect against buffer overflows.
\end{abstract}

%% file: sections/introduction.tex
\section{Introduction}
WebAssembly is quickly becoming a popular compilation target for a variety of code. It offers a fast, portable runtime and is supported by all major browsers\cite{wasm_usenix20}. However, it is highly susceptible to buffer overflow attacks.

For more than ten years, buffer overflows have been the most common security vulnerability \cite{bufferoverflow}. There exists many different methods of protections, and two common methods are are stack canaries \cite{canaries} and ASLR \cite{aslr}. For languages like C or C++ on the x86 architecture, the compiler has built in protections to prevent various buffer overflow attacks. However, the WebAssembly compiler offers no such protections. Any code that had a vulnerability would be offered some protections when compiled natively, but when compiling to WebAssembly no protections are offered.

While sandboxing does provide some protections for WebAssembly, it isn't sufficient for all use cases. The most common use case of WebAssembly is for the user to download a WebAssembly binary and run it locally in their browser. In this case, the attacker can only attack their local machine, so strong security measures aren't necessary. However, WebAssembly is beginning to be deployed on edge clouds and platforms not controlled by the attacker \cite{swivel}. In this scenario, ensuring that the code is properly protected is critical.


Previous works have dealt either with the security of the virtual environment where the binary is ran \cite{gobi, swivel} or the security of not leaking code that contains intellectual property \cite{leakprotect}. There has not been a work that proposed protections against buffer overflows.
This paper will focus on adding stack canaries and ASLR protections to WebAssembly binaries in order to protect the un-manged portions of WebAssembly's linear memory. These protections will only modify the binaries and will not rely on any modifications to the virtual machine that runs the binaries. 

Section 2 provides a more detailed look at WebAssembly and how it differs from x86, while section 3 analyzes different design options and trade offs. Section 4 delves into the attacks and the implementation for the protections. Section 5 examines the cost of the protections. Section 6 discusses the limitations of the current implementation and section 7 stipulates future ways to further improve the security of WebAssembly.


%% file: sections/background.tex
\section{Background of WebAssembly}

Since WebAssembly is still a relative new programming language, we introduce the core concepts of WebAseembly. 
\subsection{Overview}
WebAssembly \cite{zhang2021tapping, zhang2023first, feng2021allign, feng2025optimus, cheng2024efflex, liao2025catp, feng2024f3, cheng2024vetrass, han2022francis, yi2025monom, lyu2026safeguard, lyu2026educational} uses \texttt{wasm} as the binary format, and the \texttt{wasm} binary code is executed on WebAssembly's virtual machine. To make it easier to understand and debug the \texttt{wasm} binary code, WebAssembly also provides another human-readable format called \texttt{wat}. 
\begin{figure}[]
\includegraphics[width=0.45\textwidth]{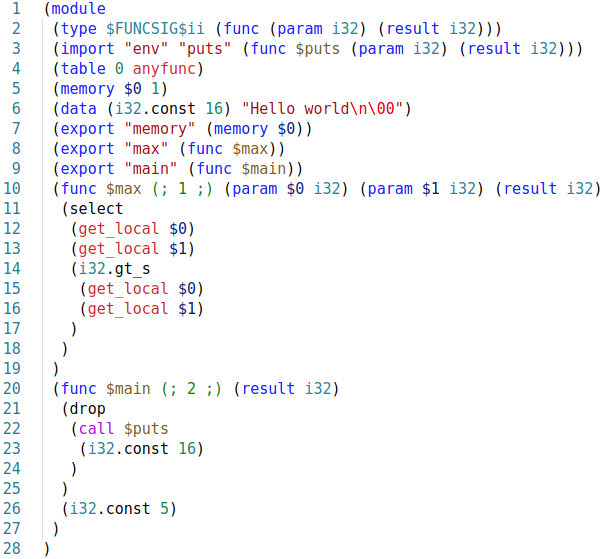}
\caption{Example of WebAssembly program in \texttt{wat} format}
\label{fig:wat}
\end{figure}

Figure \ref{fig:wat} shows a simple WebAssembly program in the \texttt{wat} format. Each \texttt{wat} file contains one module definition. Inside the module, there are declarations for functions, global variables, and at most one linear memory and indirect call table. WebAssembly provides the binary toolkit\cite{wabt_github} to help developers convert between \texttt{wasm} binary format and \texttt{wat} human-readable format. 

\subsection{Managed and Unmanaged Data}
There are two different kinds of data regions inside the WebAssembly virtual machine.

1) \textbf{Managed data}: Local variables, global variables and return address reside in dedicated data regions controlled by the WebAssembly VM.  Different from X86, there is no way to get the address of managed data. Users need to use WASM instructions like \texttt{local.get \$0} to access the managed data. This means that managed data is protected by the WebAssembly VM. 

2) \textbf{Unmanaged data}: WebAssembly only provides four primitive data types: 32-bit and 64-bit integers (\texttt{i32}, \texttt{i64}) and single and double precision floats (\texttt{f32}, \texttt{f64}). To support more complex data types in high-level language like \texttt{Array} and \texttt{Struct}, WebAssembly needs to allocate these data structures to the \textit{linear memory}.  The linear memory is a contiguous, global array of bytes, which is an unmanaged area of the WebAssembly VM. These contiguous bits are all writable. Users can use \texttt{Load} and \texttt{Store} instructions to read and write any address inside the linear memory. Similar to X86, most WebAssembly compilers places the stack onto linear memory and maintains the stack pointer as a global variable across functions. However unlike x86, the WebAssembly VM doesn't provide any security mechanisms like stack canaries or  in the linear memory. 

\subsection{Indirect table}
In a high-level language like C/C++, function pointers are one of the more important components of the language. Function pointers allow developers to do indirect function calls. To implement the function pointer, WebAssembly provides the indirect table to handle indirect calls. In the indirect table, each index is mapped to one function defined in the WebAssembly program. Users can provide the index and use the \texttt{call\_indirect} instruction to indirectly invoke the corresponding function. WebAssembly ensures the type consistency by checking the function signature before the indirect call. 

\subsection{WebAssembly Compiler}
% ...
WebAssembly is a low-level byte code, which can be used as an compilation target for high-level programming language. Currently, there are various compiler projects that can compile from high-level languages like C/C++ , Rust and Go. The most popular compiler is Emscripten\cite{emcc}, which compiles C code to WebAssembly for the browser and also adds some JavaScript code to support \texttt{printf} functions.


\section{Threat model and Design Options}
\subsection{Threat model}
In our threat model, we assume that attackers can control the input of the victim WebAssembly binary code. This means that attackers can provide malicious input to conduct stack-based buffer overflow attack proposed by previous works\cite{wasm_usenix20}. We also assume that attackers can get a local copy of the WASM binary code and do offline analysis.

\subsection{Design Options}
\label{sec:design_op}
There are two places where WASM's binary security protection can be enhanced. One option is to modify existing compilers like Emscripten that compiles to WebAssembly. This approach has larger protection space since the compiler keeps information that maps from the source code to WASM code. Another option is to build a transpiler which takes unsafe WebAssembly binary program as input and output an enhanced binary. Using the binary toolkit provided by the WebAssembly community, the transpiler is easier to implement but also has limited protection space compared to the first approach. However, it makes updating existing WebAssembly binaries easier as it doesn't require recompiling the source code. Due to the time constraint, we choose the second approach to build a secure transpiler to enhance WASM's binary security. 

%% file: sections/implementation.tex
\section{Implementation}
In this section, we describe the implementation details of the ASLR and the Stack Canary.
\subsection{Overview}
Our approach picks a lightweight solution to protect the WebAssembly code. We do not modify the original C code or the compiler. Instead, a $protector$ program is developed to inject protection code directly to the \codesm{.wasm} or \codesm{.wat} file (C assembly code alike). 

Our two protection approaches, namely ASLR and Stack Canary, are included in the $protector$.
\subsection{Random Value Generation}
To achieve strong security guarantees, we designed our random value generator to use the start time of a function execution as the seed, so that the value would be different for each function execution. This means that even if an attacker were to learn the seed of one function call, they will only be able to use that seed for a single function and all other functions will remain protected. To implement the overall workflow, we first used Emscripten to compile a C program that calls the library function $srand$ and $rand$ with the current time as the seed. Then, we studied the generated .wat file and the JavaScript file to extract code that correspond to $srand$, $rand$, and $time$. We noticed that in the .wat file, the $time$ function is imported as a function of module "env" whose implementation is in the JavaScript file. Therefore, to enable random canary value generation, our protector first introduces code corresponding to the $srand$ and $rand$ function into the .wat file. The protector then updates the JavaScript file to include the $time$ function implementation for the .wat file to import.

\subsection{ASLR} \label{subsection:ASLR}
In x86, ASLR (Address space layout randomization) randomizes the memory layout between the stack, heap, libraries, etc. By introducing ASLR, we can limit the attacker from analyzing the static memory layouts and overflowing a malicious buffer to a specific location. 

Our ASLR approach differs slightly different from x86's as it is challenging to perform the exact behavior in the WebAssembly (as described in section ~\S\ref{sec:lesson_learn}). However, our ASLR implementation is function-wise and has a better scope of protection on the attacks. We adjust the global heap pointer randomly before and after any function called instead of randomizing the location of the heap and stack pointer at the program start point. This has the benefit of protecting the callee function from the caller function. Because we added a randomized memory offsets between stack frames, the callee function cannot be sure where the parent frame resides.

\begin{lstlisting}[language=WebAssembly, caption=example of ASLR protection, label=aslrExample]
global.get $__stack_pointer
local.get 8
i32.const 26
i32.shr_s
i32.const 2
i32.shl
i32.sub
global.set $__stack_pointer
\end{lstlisting}
\subsubsection{Implementation Details}
As shown in Listing ~\ref{aslrExample}, at the beginning of each function, we generate a random value and subtract the global stack pointer by the result of shifting the generated random number right by 26 and then shifting left by 2 (to align the resulting stack pointer), and then at the end of the function we recover the global stack pointer to its original value. It is noted that the random number is generated during runtime. Therefore, the attacker cannot easily observe the number in execution. However, the ASLR approaches in WebAssembly could still suffer from frequency analysis as described in section ~\S\ref{subsec:aslr_limitation}.

\subsection{Stack Canary} \label{subsection:stackcanary}
A unique feature of WebAssembly is the separation of unmanaged linear memory (where local and global variables reside) and the managed call stack that holds return addresses. Therefore, attackers can't corrupt return addresses through a stack-based buffer overflow. This fact leads to discussion on whether stack canaries are unnecessary for WebAssembly \cite{WasmChasm}. However, as stack-based buffer overflows can still overwrite variables in the frames above the current frame on the unmanaged stack, we believe it is still worthwhile to design and implement stack canaries for WebAssembly.
\subsubsection{Random Canary Placement}

One challenge of incorporating stack canaries without modifying the compiler is determining where to store the original canary value for verification later. Since the .wat file may include instructions that use hard-coded linear memory indices (e.g., load and store), and the linear memory is vulnerable to stack-based buffer overflows, we decide to store the original canary value as a WebAssembly function local instead of a value on the linear memory. Note that as opposed to local variables that reside in frames on the linear memory, WebAssembly function locals live on the managed call stack like the return addresses. Therefore, a WebAssembly function local doesn't interfere with existing memory accesses and is not as susceptible to buffer overflow attacks (see ~\ref{subsec:aslr_limitation}). 

The concrete workflow of placing the stack canaries and storing the original values is as follows: we append the $rand\_canary$ function that takes the current time as input and returns a random value to the end of the .wat file; next, at the very beginning of each function in the .wat file, we inject a piece of code that first gets a random canary value through calling the imported $time$ function and then appended $rand\_canary$, and then store the canary value in a local variable; finally, the injected code subtracts the global stack pointer by 4 and stores the canary value at the resulting stack pointer position, and then stores the address of the stack canary (i.e., the new stack pointer position) in another local. An example of this injected prologue is shown in Listing ~\ref{canaryPrologue}.

\begin{lstlisting}[language=WebAssembly, caption=example of stack canary protection prologue, label=canaryPrologue]
i32.const 0
local.set 7
local.get 7
call $get_time
local.set 7
local.get 7
call $rand_canary
local.set 7
global.get $__stack_pointer
i32.const 4
i32.sub
local.set 8
local.get 8
local.get 7
i32.store
local.get 8
global.set $__stack_pointer
\end{lstlisting}
\subsubsection{Canary Verification}
Immediately before a function returns, we verify that the canary on the stack matches its original value stored as a local. Before the "return" instruction or the last parentheses of the function, we inject a piece of code that loads the current value at the stack canary address and compares it with the original canary value. If the two values match, then it increments the global stack pointer by 4 (recall that at the beginning of the function we subtract the stack pointer by 4 to allocate space for the canary). Otherwise, the function will encounter an "unreachable" instruction, which leads to a runtime error. An example of this injected epilogue is shown in Listing ~\ref{canaryEpilogue}.
\begin{lstlisting}[language=WebAssembly, caption=example of stack canary protection epilogue, label=canaryEpilogue]
block
local.get 8
i32.load
local.get 7
i32.eq
br_if 0
unreachable
end
global.get $__stack_pointer
i32.const 4
i32.add
global.set $__stack_pointer
\end{lstlisting}

%% file: sections/evaluation.tex
\section{Evaluation}

In this section, we evaluate the cost and effectiveness of our ASLR and Stack Canary protections. 

\subsection{Protection Overhead}
ASLR introduces extra instructions before and after each function call, while stack canary adds instructions at the beginning and end of function declarations. We summarized the number of different types of instructions added by introducing our two approaches in Table ~\ref{table:function_inst_overhead}. The types of instructions we use are arithmetic instructions (i.e., const, add, and sub), variable instructions (i.e., local.set, local.get, global.set, and global.get), memory instructions (i.e., load and store), and control instructions (i.e., unreachable, br\_if, return, call, and block \{instructions\} end).

To evaluate the protection overhead of one specific program $p$, the total number of extra instructions that need to be executed can be calculated as following. 

\begin{equation}
    Overhead_{ASLR}(p) = overhead_{ASLR} \times \#function\_calls
\end{equation}

\begin{equation}
    Overhead_{Canary}(p) = overhead_{Canary} \times \#function\_calls
\end{equation}

The $overhead_{ASLR}$ and $overhead_{Canary}$ as the number of instructions needed to be add for each function, listed in the Table \ref{table:function_inst_overhead}.

\begin{table}[!ht]
    \renewcommand{\arraystretch}{1.3}
    \centering
    \begin{tabular} { |c|c|c| }
    \hline
    Instruction Type & ASLR & Stack Canary \\
    \Xhline{3\arrayrulewidth}
    Arithmetic instructions & 21 & 16 \\
    \hline
    Variable instructions & 12 & 16 \\
    \hline
    Memory instructions & 0 & 2 \\
    \hline
    Control instructions & 0 & 5 \\
    \hline
    \end{tabular}
    \vspace{10pt}
    \caption{Instruction overhead in a function}
    \label{table:function_inst_overhead}
\end{table}

When we enable both ASLR and Stack Canary, the protector only inserts the series of instructions for generating a random value (e.g., the first eight instructions in Listing \ref{canaryPrologue}) once. Therefore, the protection's overhead when using both mechanisms is smaller than the sum of $overhead\_ASLR$ and $overhead\_Canary$.

As we are protecting every single function (except the $stackAlloc$ function which modifies and returns the value of the global stack pointer, and hence interferes with our protection), the overhead the protections introduce vary greatly with the program being protected. Thus it is difficult to know exactly how much overhead this protection introduces. Further analysis will have to be completed.


\subsection{Protection Coverage} \label{subsec:protection_coverage}
The major attack primitive we target is the stack-based buffer overflow attack primitive extensively discussed in previous works \cite{wasm_usenix20, Bergbom, WasmChasm}. As highlighted in \cite{wasm_usenix20}, among the 98,924 functions from 26 WebAssembly binaries the authors analyzed, one third (32,651) stored data on the unmanaged stack. Therefore, there is a high chance that the attackers will find some overwritable variables on the stack. Moreover, the authors also found that averaged over the 26 programs, 9.8\% of call instructions are indirect calls, and 49.2\% of functions are a valid target of at least one indirect call in the same program \cite{wasm_usenix20}. So, the risk of control flow hijacking through buffer overflow is high. As our two protection mechanisms can prevent buffer overflow attacks to different extents, we consider our protector as effective in handling a major vulnerability of WebAssembly programs.

In the following section, we pick one public released vulnerable program \cite{wabt_github} provided in \cite{wasm_usenix20} to validate the effectiveness of our protection approaches. More details of this experiment and the procedures to reproduce can be found in Appendix ~\S\ref{sec:appendix}.

\subsubsection{ASLR Protection}
To test our ASLR protection, we use the vulnerable program mentioned above with malicious input to trigger the evil function call by overflowing the buffer in the callee function. With our ASLR protection, the evil function is no longer called because we randomized the offsets between each caller and callee function. However, the robustness of our protection depends on the range of random values chosen. More details about the limitation of this approach can be found in section ~\S\ref{subsec:aslr_limitation}.

\subsubsection{Stack Canary Protection}
As discussed in section \S\ref{subsection:stackcanary}, we place the stack canary between the current frame and the frames above on the unmanaged stack. Thus, any attempt to overwrite variables in other frames through buffer overflow will corrupt the canary value, leading to a canary verification failure and a runtime error. We applied our stack canary protection to the stack-based buffer overflow attack primitive demonstration program in \cite{wasm_usenix20}. We successfully confirmed that our protector's output program crashes due to canary value mismatch when the input is long enough to overwrite variables in other frames.

%% file: sections/limitation.tex
\section{Discussion}
\subsection{Limitations}

One problem that plagues both protection methods is the fact that global variables can be overflowed\cite{wasm_usenix20}. This is problematic because global variables are stored in the manged memory, so overflowing them could overwrite any of return addresses or canary values we store. However, without modifying the WebAssembly VM, there is no way to protect against this. In languages like C/C++, there would exist an unwritable page and if a buffer overflow tried writing to the page an error would be thrown. This can't be done with WebAssembly. If we were to modify the VM, then we could add protections. 

\subsubsection{ASLR} \label{subsec:aslr_limitation}
The limitation of ASLR in WebAssembly is previously discussed in ~\cite{wasm_usenix20}. The 32-bit address is used in WebAssembly to locate the linear memory, which does not have enough random space to prevent the attacker from launching a brute-force-like attack to guess the random number. 

Our function-wise ASLR also holds the same limitation. Our approach adjusts the global stack pointer by a generated random number. The range of this number can only be in a certain range, or otherwise, the program exceeds the memory bounds. Our current implementation uses \texttt{i32.shl} and \texttt{i32.shr} instructions to round the random number in the range of [0, 256) to satisfy the constraint of alignment and limited stack space. So it could be brute forced in about 64 tries. 
\subsubsection{Stack Canary}
Since we place the stack canary above the current stack frame's local variables, we can't detect buffer overflow attacks that overwrite local variables while not touching the stack canary. Therefore, attackers are still capable of hijacking the control flow by corrupting function pointers. Stronger stack canary implementations handle this by reordering variables to place all buffers in higher addresses and non-buffer variables in lower addresses \cite{Richarte02}. We believe such reordering is also applicable to WebAssembly, but would require some forms of compiler modification.

Another limitation with our stack canary implementation is that it lacks meaningful error messages. Currently, when a canary verification fails, the program simply crashes with an unrecoverable runtime error thrown. In the future, we would like to introduce more informative error messages that explicitly pinpoint the cause and the vulnerable function.

Our implementation generates a new random number every time a function is called as we use the time as the random seed. This could be problematic as if the attacker can control when a function is called, they could get the random seed and thus extract the random number generated. Thus they could get the canary value and overwrite it without being detected. To prevent this, we would have to change the way we generate random numbers. Another way would be to generate some magic value at runtime and use this constant value for all canary values. However, this doesn't offer as much protection, especially since a control flow attack could be used to glean this magic value (see \ref{subsec:aslr_limitation} to see why we can't protect against control flow attacks).

\subsection{Lessons Learned and Future Work}
\label{sec:lesson_learn}
\subsubsection{Binary-level insertion}
Our $protector$ program can only support text-level code injection at this stage. We need to convert the WebAssembly binary file to text file before code injection and convert it back eventually. This is not a huge cost but we expect to inject the code in a more generic and easy way in the future, like inserting the binary code directly.

\subsubsection{Indirect Table Protection}
As shown in previous work\cite{wasm_usenix20}, the indirect table is widely used in many WebAssembly programs. Since the indirect table itself is a simple mapping from index to the functions, it is often the attackers' target to conduct control flow-based attack.

We have some initial attempts to randomize the the layout of element table at compile time and attackers can't know the index of target function. But the weakness of this defense strategy is that there are only limited functions in the indirect table. To ensure the type correctness, we can only permute functions with the same signature, which further reduce the number of possible permutations. So if attackers have enough attempts, they can still figure out the index of the victim function. Furthermore, we initially assumed that attackers would have the exact binary being used. Since the permutation only occurs once at compile time, the attacker would be able to see the permutation we used, thus rendering the protection ineffective. In our next step, we try to evaluate the possibility of adding run-time protection for the indirect table. 
\subsubsection{Full Compiler Implementation}
Since our \texttt{protector} program is separate from the Emcripten compiler, it doesn't have any information like a symbol table that can be used to do deeper analysis of the WASM binary code. In previous work \cite{wasm_usenix20}, they also show that current Emscripton provides no protection between two buffers declared in the same function. So it's possible to overflow one buffer to modify the contents of another buffer. However, our \texttt{protector} has no information about the address of these two buffers and cannot put cookies between them to protect. As discussed in the section \ref{sec:design_op}, adding defense mechanisms directly into the compiler will provide much larger protection space. 

\section{Conclusion}

We have proposed two protection methods for buffer overflows in WebAssembly. These protections are executed on a per function basis, which improves security isolation. While these protections protect the caller frame from the callee frame, there are still many problems faced when doing these protections through a transpiler. In future work, we would consider both adding security protections in the compiler from a high level language to WebAssembly and in the WebAssembly VM itself. WebAssembly's growth shows how powerful of a tool it can be, but as it becomes more powerful and prominent, it needs to have better security. This is just the first step in doing so.

%% file: sections/appendix.tex
\section{Appendix} 
\label{sec:appendix}
\subsection{Environment}
The tools \codesm{emcc} and \codesm{wabt} need to be installed. The \codesm{emcc} version we used is 2.0.34 and the \codesm{wabt} version is 1.024. Our $protector$ relies on Python3. These tools are supported across MacOS, Linux and Windows.
\subsection{How to Run}
Several Makefile rules are provided for reproducing our experiment. 
\begin{itemize}
  \item \codesm{\$ make without\_protection},
  \item \codesm{\$ make with\_ASLR}, 
  \item \codesm{\$ make with\_canary\_ASLR},
  \item \codesm{\$ make with\_canary}
\end{itemize}

\subsection{Codebase Structure}
\begin{itemize}
  \item \codesm{\codesm{protector.py}}

  The $protector$ as paper described.
 
  \codesm{Usage: python3 ./protector [name].wat \\ -[ASLR/canary/canary\_and\_ASLR]}
      
  \item \codesm{main.c}

  The vulnerable program used in section ~\S\ref{subsec:protection_coverage}
 
  \item  \codesm{canary\_insertion.txt}
  
  The helper function inserted to help Stack Canary protection.
  
  \item \codesm{output\_examples}
  
  The pre-executed output examples for the 4 Makefile rules above.
\end{itemize}

